\documentstyle[11pt,newpasp,twoside,epsf]{article}
\begin{document}
 
\title{Contribution of the VLBA Network to Geodynamics}
  \author{Leonid Petrov}
  \affil{NVI, Inc./NASA GSFC, Code 926, Greenbelt, 20771 MD, USA}
\begin{abstract}
  A decade of observations in geodetic mode with the VLBA network gave
valuable results. Approximately 1/5 of all geodetic observations are
recorded in VLBA mode and processed at the Socorro correlator.
Ten years of observations allowed us to reliably measure slow intra-plate
motion of the VLBA stations located on the North-American plate. It also
helped us to achieve scientific objectives of geodetic programs: verify models
of harmonic and anharmonic site position variations caused by various
loadings, improve models of core-mantle boundary, investigate mantle rheology
and solve other tasks.
\end{abstract}
 
\section{Introduction}
 
  Method of VLBI first proposed by Matveenko, Kardashev and Sholomitsky (1965)
is used for two major applications: making images of compact radio sources
and measuring antenna's motion with respect to extragalactic source. Fringe
amplitudes and phases measured with precision  $10^{-1}$--$10^{-3}$ are used
as primary observables for imaging; group delays measured with relative
precision $10^{-8}$--$10^{-10}$ are used as observables for geodynamics
applications. Group delay is sensitive to the projection of the baseline vector
to the source direction. Having observed many sources in different directions
and different baselines we are able to solve for source positions, site
positions, Earth orientation parameters (EOP) and the rate of their changes
in a least square solution. The paper of Overs, Fanselow and Jacobs (1998)
provides an overview of geodetic VLBI technique.
 
\section{Precision and accuracy}
 
  There are several measures of how precise results are. Formal uncertainties,
which are calculated according to a law of propagation of errors of an
individual observation, are the first measure. This estimate gives us an upper
limit of precision, since an estimation model usually does not account
correlations correctly. Another measure of precision is repeatability --- the
root mean square deviation of the series of estimates of the targeted parameter
with respect to the average. And the last measure is accuracy: the measure of
how far our estimate of the targeted parameter is from the true value.
Table~\ref{t:acc} summarizes the current level of precision and accuracy
of geodetic VLBI.
 
\begin{table}
  \caption{Precision and accuracy of VLBI.}
  \begin{center}
  \begin{tabular}{ |l|l| r@{\quad}r @{\quad}|@{\quad} r@{\quad}r | }
     \hline
                   &  \multicolumn{1}{c|}{EOP} &  \multicolumn{4}{c|}{Station coordinates
                                                   \rule[-1.5ex]{0.0ex}{4.0ex}} \\
     \cline{2-6} \rule[-1.2ex]{0.0ex}{4.0ex}
                   &                &  \multicolumn{2}{|c}{One day}    &
                                       \multicolumn{2}{c|}{All data }  \\
     \cline{3-6} \rule[-1.2ex]{0.0ex}{4.0ex}
                   &          & vert    & horiz  & vert   & horiz             \\
     \hline
     Precision     & 0.3 nrad & 2.0 mm  & 1.5 mm & 0.7 mm & 0.2 mm \rule[-1.0ex]{0.0ex}{4.0ex} \\
     Repeatability & 0.5 nrad & ---     & ---    &   8 mm &   2 mm \rule[-1.0ex]{0.0ex}{4.0ex} \\
     Accuracy      & 0.7 nrad & 10  mm  & 3 mm   &  10 mm &   3 mm \rule[-1.0ex]{0.0ex}{4.0ex} \\
     \hline
  \end{tabular}
  \label{t:acc}
  \end{center}
  \par\vspace{1ex}\par
  \begin{center}
  \begin{tabular}{ |l| r@{\quad}r  @{\quad}|@{\quad}  r@{\quad}r | }
     \hline
      \multicolumn{5}{|c|}{Amplitude of coherent signal in site coordinates} \rule[-1.5ex]{0.0ex}{4.0ex}\\
     \hline \rule[-1.2ex]{0.0ex}{4.0ex}
     \hspace{18.5mm} &  \multicolumn{2}{c}{Periods $> 30^d$} &  \multicolumn{2}{c|}{Periods $< 1^d\!.2$}
                                                                              \\
     \cline{2-5}
               & vert    & horiz      &  vert      & horiz       \rule[-1.3ex]{0.0ex}{4.0ex}   \\
     \hline
     Precision & 1--4 mm & 0.2--0.7 mm & 0.7--1.5 mm & 0.2--0.4 mm \rule[-1.0ex]{0.0ex}{4.0ex} \\
     Accuracy  & 2--5 mm & 0.4--1.2 mm & 1.0--2.0 mm & 0.3--0.6 mm \rule[-1.0ex]{0.0ex}{4.0ex} \\
     \hline
  \end{tabular}
  \end{center}
  \par\vspace{-6mm}\par
\end{table}
 
\section{Scientific objectives}
 
  Although site positions and source coordinates are determined in processing
VLBI observations, and they can be useful for particular applications,
such as source mapping, determination of {\em positions} is not the objective
per se. The objective of observations is to measure {\em changes} in site
positions and source coordinates. Investigation of changes in positions
of the sites allows us to make judgments about the forces which caused the
motion and the response the Earth to external forcing.
 
  When the whole VLBI dataset is analyzed, the following parameters are
estimated:
\begin{itemize}
    \item Secular site velocities. Causes: 1)~plate tectonic;
          2)~po-seismic slips; 3)~post-glacial isostatical adjustment;
          4)~environmental changes 5)~??
 
    \item amplitudes of harmonic site position variations. Causes:
          1)~solid Earth tides; 2)~ocean tides; 3)~??
 
    \item anharmonic site position variations. Causes:
          1)~atmospheric pressure loading; 2)~hydrology loading;
          3)~non-tidal ocean loading; 4)~??
 
    \item variations in the Earth orientation parameters. Causes:
          1)~torques exerted by the Sun and planets; 2)~exchange of angular
             momentum with the atmosphere; 3)~exchange of angular momentum
             with the ocean; 4)~earthquakes? 5)~??
\end{itemize}
 
  One of the examples of achieving these objectives are results of analysis
of the amplitudes of the Earth's nutation fitted to VLBI data.
 
\begin{figure}[h]
  \caption{Real and imaginary part of the observed nutation transfer function.
           Solid line corresponds to the case of rigid Earth.}
  \par\vspace{2ex}\par
  \epsfclipon
  \plottwo{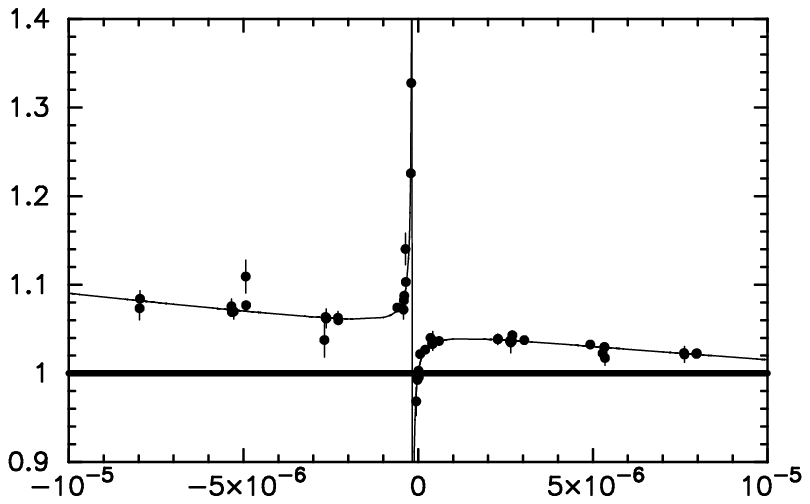}{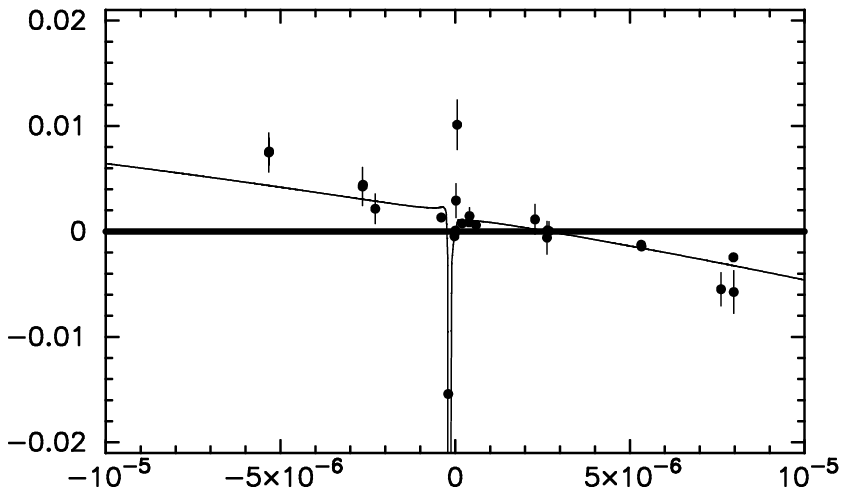}
  \par
  {\hspace{26mm} \small Frequency, rad/sec}
  {\hspace{33mm} \small Frequency, rad/sec}
  \par\vspace{-2ex}\par
\end{figure}
 
  In the case if the Earth responses to external torques as an absolutely
rigid body, the complex transfer function would be (1,0). The presence of the
liquid core causes the resonance at the retrograde free core nutation
frequency. Dissipation in the ocean and in the liquid body causes appearance
of the non-zero imaginary part of the transfer function. Fitting empirical
transfer function to a model allowed Buffett, Mathews, \& Herring, (2002) to
estimate so-called basic Earth paramenters which includes the core mantle
boundary flattening and the magnitude of the magnetic field at the core mantle
boundary (0.7mT).
 
\begin{figure}[h]
  \par\vspace{-3ex}\par
  \caption{Observed (disks) and theoretical (circles) amplitudes of the Up
           component of the semi-diurnal ocean loading at station MK-VLBA,
           in-phase (left) and out-of-phase(right), in mm.}
  \epsfclipon
  \epsfxsize=120mm
  \epsffile{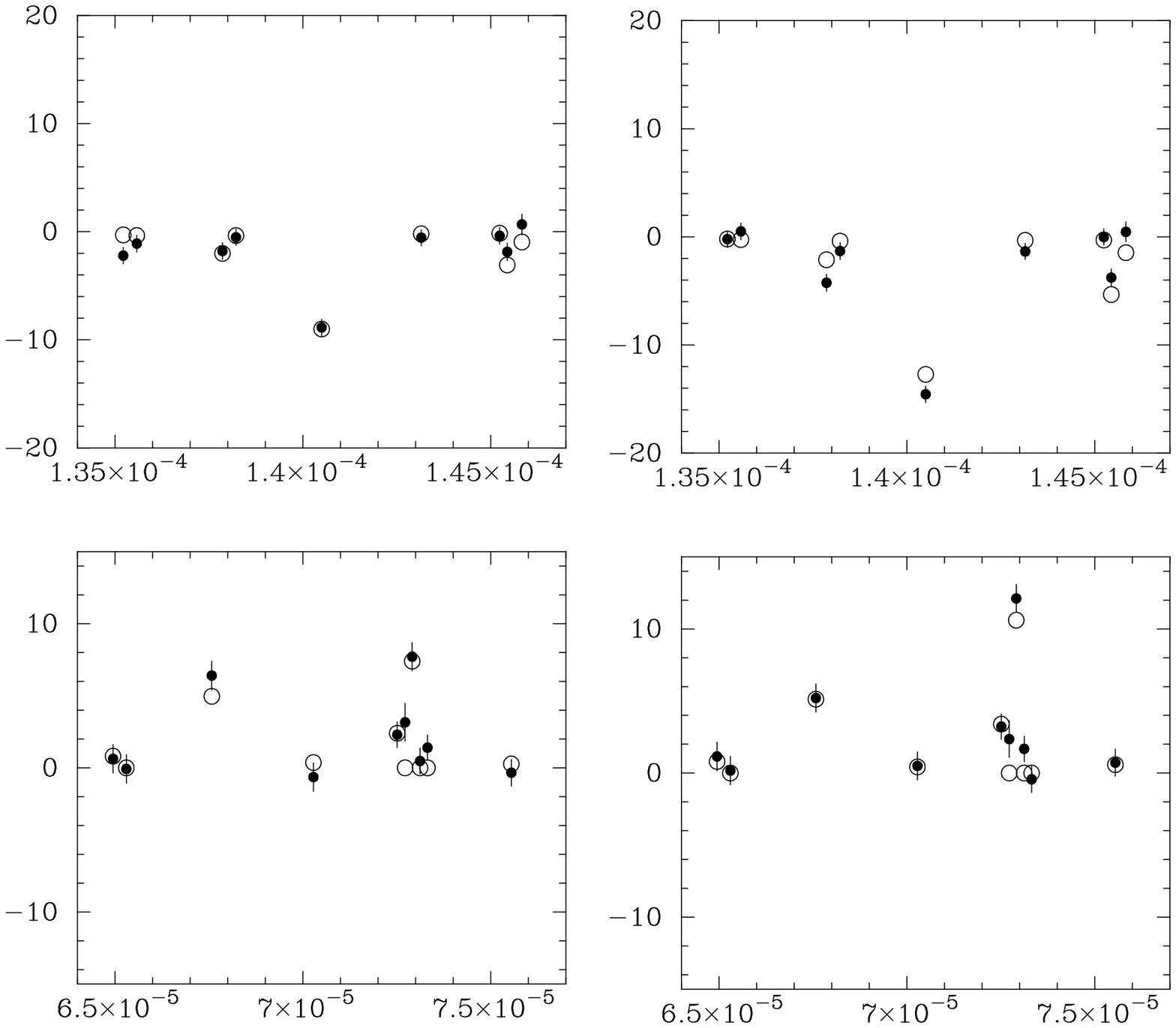}
  {\hspace{28mm} \small Frequency, rad/sec}
  {\hspace{30mm} \small Frequency, rad/sec}
  \label{f:oclo}
  \par\vspace{-3ex}\par
\end{figure}

  VLBI observations allowed us to detect a secondary tidal effect: site
displacements caused by water mass redistribution due to ocean tides.
Mass loading results in crust deformation even thousands kilometers
away from the mass load. It was found that observed amplitudes are in a rather
good agreement with models (fig.~\ref{f:oclo}. Comparison of the observed
amplitudes with theoretical amplitudes at many VLBI sites, including VLBA
antennas, allowed Petrov and Ma (2003) to discriminate between competing
models of ocean tides. It is relatively easy to validate ocean tides models
close to the shore where tide gauges are located, but it is a challenge to
validate ocean tides models at open ocean.
 
\begin{figure}[h]
  \caption{Horizontal displacement due to atmosphere pressure loading
           at station Pietown and vertical displacement at station BR-VLBA
           in mm.}
  \par\vspace{1ex}\par
  \epsfclipon
 
  \mbox{ \epsfxsize=53mm \epsffile{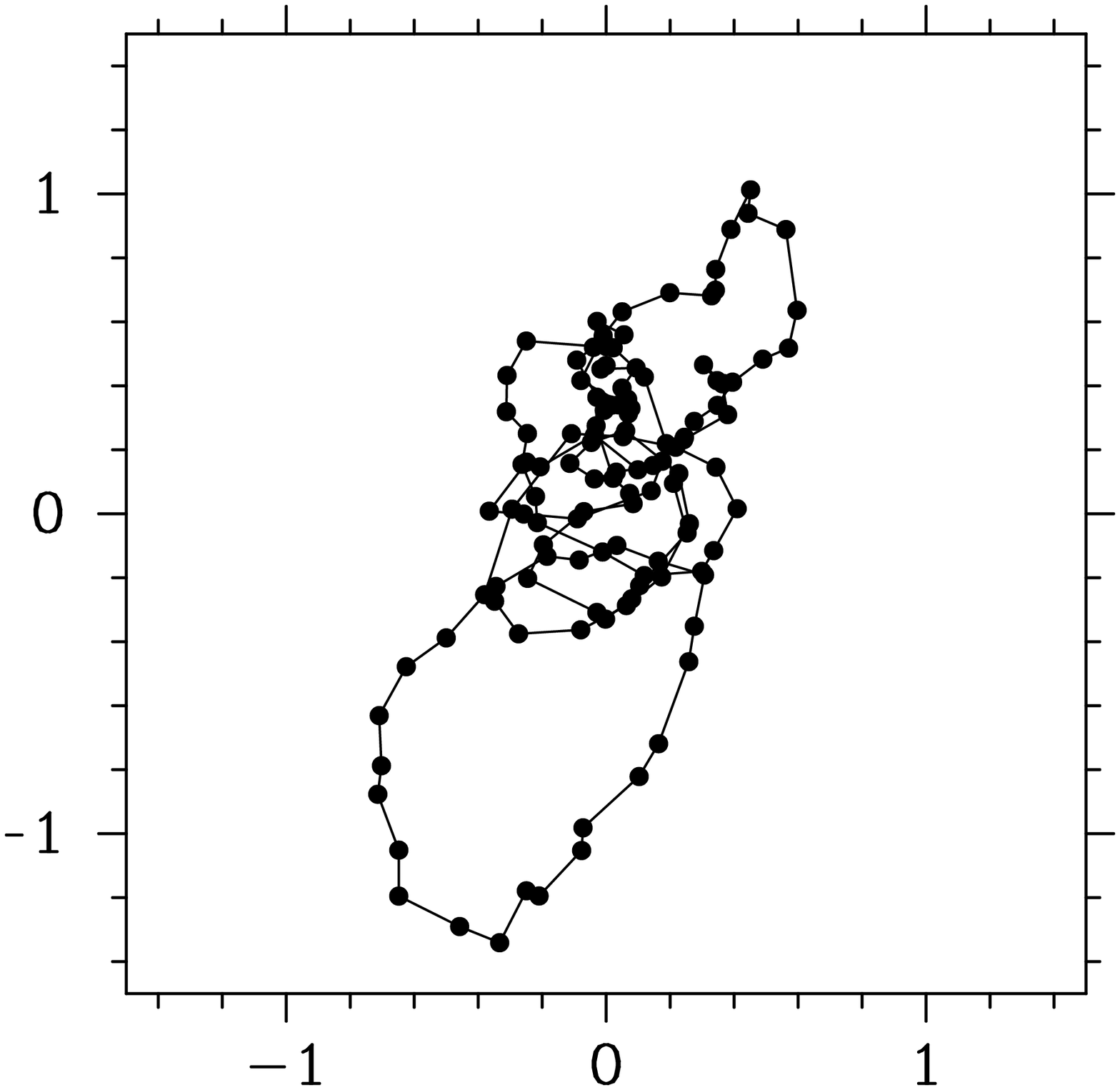} \hspace{1.5mm}
         \vbox{ \epsfxsize=75mm \epsffile{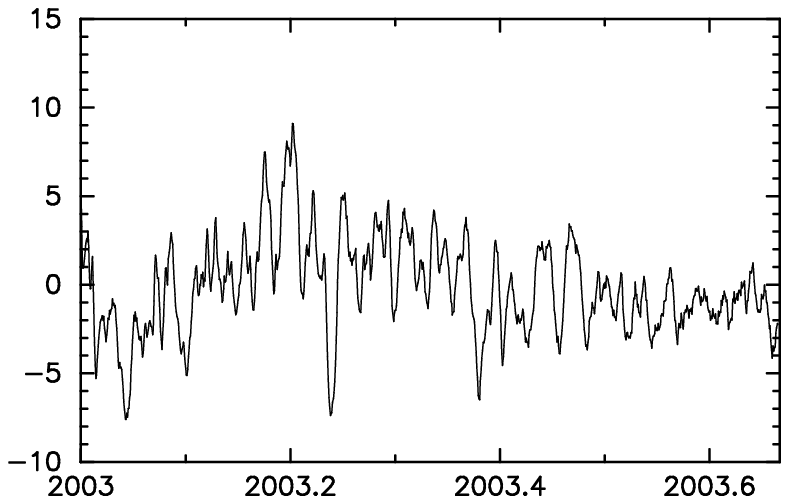} \par\vspace{1mm} }
       }
  \label{f:aplo}
  \par\vspace{-13mm}\par
\end{figure}

  Similarly to mass loading, changes of atmospheric mass above the surface
cause crust deformation due to pressure loading. Unlike tides, changes in
atmospheric pressure are not predictable. However, global numerical weather
models based on past observations of the world meteorological network allow
us to compute displacements due to atmospheric pressure loading with latency
several days. Examples of the atmospheric pressure loading are presented in
fig.~\ref{f:aplo}. Comparison of the admittance factors of the theoretical
time series for all sites, including VLBA sites, showed surprisingly well
agreement between the theoretical model and observations, even for horizontal
loading which does not exceed 1~mm for most of the sites. This allowed Petrov
and Boy (2003) to set the upper limit (4\%) of possible errors of loading
Love numbers which describe elastic properties of the Earth.
 
\section{Contribution of VLBA}
 
  VLBA stations started to participate in geodetic program even before
construction of the VLBA network has been completed. By June 2003 the total
number of observations used in analysis was \mbox{\it 4\,032\,371}. The number
of observations with at least one VLBA site was \mbox{$\it 894\,690 $}. Thus,
approximately 1/5 of geodetic observations are done at the VLBA network.
 
\begin{table}
  \caption{VLBA station velocities with respect to the rigid North American
           plate. Unit: $\beta=v/c$ \quad
           ( 1 mm/yr $ \approx 1.06 \cdot 10^{-19} $ v/c ). Last column
           indicates whether the station was used as defining for horizontal
           or vertical motion of the plate.}
  \par\vspace{0.5ex}\par
  \begin{tabular}{|l| r@{\,}r@{\,}c@{\,}r | r@{\,}r@{\,}c@{\,}r |
                    r@{\,}r@{\,}c@{\,}r |l|}
     \hline
     Station   &  \multicolumn{4}{c|}{Up} &  \multicolumn{4}{c|}{East} &  \multicolumn{4}{c|}{North} &
               Def \\
     \hline
     \sc br-vlba & (&  0.9 &$\pm$& 2.5)$\cdot 10^{-19}$ & (&      2.4  &
 $\pm$&  0.7)$\cdot 10^{-19}$ & (& -0.4 &$\pm$&  0.5)$\cdot 10^{-19}$  &      \\
     \sc fd-vlba & (&  2.2 &$\pm$& 1.2)$\cdot 10^{-19}$ & (&      0.3  &
 $\pm$&  0.2)$\cdot 10^{-19}$ & (&  0.0 &$\pm$&  0.2)$\cdot 10^{-19}$  &  h   \\
     \sc hn-vlba & (&  0.0 &$\pm$& 0.1)$\cdot 10^{-19}$ & (&      0.0  &
 $\pm$&  0.2)$\cdot 10^{-19}$ & (&  0.0 &$\pm$&  0.1)$\cdot 10^{-19}$  &  hv  \\
     \sc kp-vlba & (&  2.8 &$\pm$& 1.3)$\cdot 10^{-19}$ & (&     -0.1  &
 $\pm$&  0.2)$\cdot 10^{-19}$ & (&  0.1 &$\pm$&  0.2)$\cdot 10^{-19}$  &  h   \\
     \sc la-vlba & (&  1.5 &$\pm$& 1.1)$\cdot 10^{-19}$ & (&     -0.1  &
 $\pm$&  0.2)$\cdot 10^{-19}$ & (&  0.0 &$\pm$&  0.3)$\cdot 10^{-19}$  &  h   \\
     \sc mk-vlba & (&  5.0 &$\pm$& 3.2)$\cdot 10^{-19}$ & (&\bf -57.2   &
 $\pm$&  1.6)$\cdot 10^{-19}$ & (&\bf 57.7 &$\pm$&  3.4)$\cdot 10^{-19}$    & \\
     \sc nl-vlba & (& -1.1 &$\pm$& 0.8)$\cdot 10^{-19}$ & (&     -0.1  &
 $\pm$&  0.2)$\cdot 10^{-19}$ & (& -0.1 &$\pm$&  0.2)$\cdot 10^{-19}$  &  h   \\
     \sc ov-vlba & (&  2.7 &$\pm$& 1.7)$\cdot 10^{-19}$ & (&\bf  -6.1   &
 $\pm$&  0.5)$\cdot 10^{-19}$ & (&\bf  5.4 &$\pm$&  0.5)$\cdot 10^{-19}$    & \\
     \sc pietown & (&  2.5 &$\pm$& 1.2)$\cdot 10^{-19}$ & (&     -1.2  &
 $\pm$&  0.3)$\cdot 10^{-19}$ & (& -2.1 &$\pm$&  0.3)$\cdot 10^{-19}$       & \\
     \sc sc-vlba & (& -3.8 &$\pm$& 3.4)$\cdot 10^{-19}$ & (&\bf  20.2   &
 $\pm$&  1.3)$\cdot 10^{-19}$ & (&\bf  5.0 &$\pm$&  0.6)$\cdot 10^{-19}$    & \\
     \hline
  \end{tabular}
  \label{t:sec_vel}
  \par\vspace{-2ex}\par
\end{table}
 
  Results of estimation of secular station velocity are presented in
table~\ref{t:sec_vel}. The station of MK-VLBA belongs to the Pacific tectonic
plate, SC-VLBA belongs to the Caribbean plate. They move rapidly
(in geophysical meaning of this word) with respect to the North American plate
where 8 other stations are located. The station of OV-VLBA is located rather
close to the plate boundary and this station, and in a lesser degree BR-VLBA,
are affected by the plate deformation near the edge of the plate. Peculiar
horizontal velocity of Pietown is not yet completely understood.
As Gordon (2003) indicated, it may be related to antenna's tilting. Six other
stations do not exhibit relative horizontal motion with respect to each other
at the level exceeding 1 mm during entire history of the VLBA.
 
  Exceptionally high precision of the Earth orientation parameters makes
20~station experiments correlated at the VLBA correlator especially
attractive for investigation of free core nutation -- a poorly understood
phenomenon of variable amplitude with the period around 430 days. Unlike forced
nutation caused by torques exerted by Moon, Sun and planets, the mechanism
of excitation of the free core nutation is not known. Currently it cannot
be predicted and should be monitored like polar motion.
 
\begin{figure}[h]
   \par\vspace{-2ex}\par
   \caption{Observed free core nutation after 2001.0 from 20 station
            VLBA experiments in nrad.}
   \par\vspace{1.0ex}\par
   \epsfxsize=48mm
   \epsfclipon
   \mbox{ \hspace{40mm} \epsffile{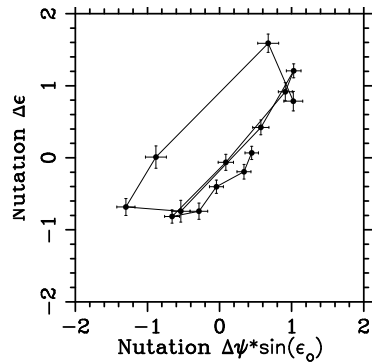} }
   \par\vspace{-8ex}\par
\end{figure}

\end{document}